\documentclass[aps,prb]{revtex4}

\usepackage{graphicx}
\usepackage{amssymb}

\begin{document}


\title{Sub-diffraction light propagation in fibers with anisotropic dielectric cores}
\author{Alexander A. Govyadinov}
\affiliation{Physics Department, Oregon State University, 301 Weniger Hall, Corvallis, OR 97331, USA}
\author{Viktor A. Podolskiy}
\email{viktor.podolskiy@physics.oregonstate.edu}
\affiliation{Physics Department, Oregon State University, 301 Weniger Hall, Corvallis, OR 97331, USA}

\begin{abstract}
We present a detailed study of light propagation in waveguides with anisotropic metamaterial cores. We demonstrate that in contrast to conventional optical fibers, our structures support free-space-like propagating modes even when the waveguide radius is much smaller than the wavelength. We develop analytical formalism to describe mode structure and propagation in strongly anisotropic systems and study the effects related to waveguide boundaries and material composition.
\end{abstract}

\maketitle

\section{Introduction}
Guided propagation of optical signals in isotropic dielectric fibers is possible only when the light-transmitting region of a fiber is at least of the order of the free-space wavelength\cite{landauECM}. This fact strongly limits resolution of modern optical microscopy and spectroscopy, prevents the construction of compact all-optical processing units and further development in other areas of photonics\cite{natureSNOM,sciNF,kivsharTrans,optTransOL,walba,lipsonNat}. Although it is possible to propagate $GHz$ radiation in deep subwavelength areas in coaxial cables or anisotropic magnetic systems\cite{keilman02,belovPRE,marques,hrabar}, the direct scale-down of these techniques to optical or IR domains is problematic\cite{sppHard}.

Until recently, all designs involving optical light transport in subwavelength areas relied on the excitation of a surface wave -- a special kind of electromagnetic wave propagating at the boundary between the materials with positive and negative dielectric constants\cite{atwaterNature,BozhevolnyiPRL,stockmanPRL,JoannopoulosSPP}. The spatial structure of surface waves, however, is fundamentally different from the one of ``volume'' fiber modes or free-space radiation. While it is possible to couple the radiation between the volume and surface modes, such a coupling is typically associated with substantial scattering losses and involves materials and devices of substantial (in comparison with optical wavelength) size\cite{novotny,landauECM}.

A new approach to guide {\it volume} modes in subwavelength areas has been recently introduced in Ref.\cite{funnels05}. It has been demonstrated that the waveguide with strongly anisotropic dielectric core supports propagating modes even when its radius is much smaller than the wavelength. The phase velocity of these propagating modes can be either positive or negative depending on the waveguide core material. It has been also shown that anisotropic core waveguides can be tapered and used to effectively transfer free-space radiation to and from nanoscale. These tapered systems, {\it photonic funnels}, are in a sense variable-index ($n\lessgtr 0$), volume mode-analog of the adiabatic surface mode compressor, proposed in Ref.\cite{stockmanPRL}. 

In this manuscript we describe the perspectives of light propagation in anisotropic waveguides. In Section 2 we describe the physics behind the volume mode propagation in subwavelength waveguides with perfectly conducting metallic walls. Effects of the core micro-structure on the light propagation are considered in Section 3. Section 4 is devoted to the effect of waveguide walls on mode structure, confinement, and propagation constants. Section 5 concludes the paper. 

\section{Light transmission in waveguides with perfectly conducting walls}

\begin{figure}
\centerline{\includegraphics[width=10cm]{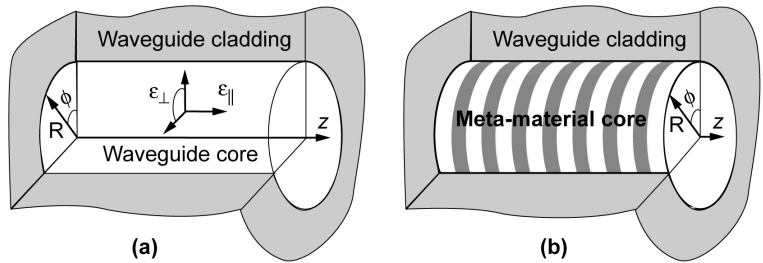}}
\caption{Schematic geometry of a waveguide with strongly anisotropic dielectric core (a). Nano-layered metal-dielectric metamaterial realization of a strongly anisotropic core is shown in (b). Optical axis and the direction of light propagation are parallel to $z$ axis.}
\label{figConfig}
\end{figure}

The geometry of the light transmission considered in this work is schematically shown in Fig.~\ref{figConfig}. The direction of light propagation is assumed to coincide with $z$ axis of cylindrical coordinate system. The optical axis of the anisotropic core material is assumed to be parallel to $z$ direction; $\epsilon_\|$ and $\epsilon_\perp$ denote the values of the effective dielectric permittivity parallel and perpendicular to optical axis respectively. 

Following the formalism, introduced in \cite{podolskiyPRB,funnels05}, the waves propagating in a waveguide with perfectly conducting walls can be represented as a series of ($TE$) and ($TM$) modes. Each mode has its own spatial structure across the waveguide, determined by a structural parameter $\kappa$ as described below. The propagating component of the mode's wavevector can be related to its frequency through the free-space-like dispersion relation: 
\begin{equation}
\label{eqnDEq}
k_z^2=\epsilon\nu\frac{\omega^2}{c^2},
\end{equation}
where $\omega$ is the angular frequency of light, $c$ is the speed of light in the vacuum, and the polarization-dependent propagation constants\footnote{ Note the difference between the geometry considered in this work (and in \cite{funnels05}) and the one in \cite{podolskiyPRB}}
\begin{eqnarray}
\epsilon_{TE}=\epsilon_{TM}=\epsilon_\perp, \nonumber
\\ \nu_{TE}=1-\frac{\kappa^2 c^2}{\epsilon_\perp \omega^2}, 
\label{eqEpsNu}
\\ \nu_{TM}=1-\frac{\kappa^2 c^2}{\epsilon_\| \omega^2}. \nonumber
\end{eqnarray}

For $TM$ waves the structural parameter $\kappa$ and the mode structure across the waveguide can be related to the $z$ component of $E$ field, which is in turn is determined from the differential equation 
\begin{equation}
\label{eqKappa}
\Delta_2 E_z+\kappa^2 E_z=0,
\end{equation}
with $\Delta_2$ being the $2D$ Laplacian operator, and an additional condition that the electric field $E_z$ satisfies the boundary conditions along the core-cladding interface\cite{landauECM}. Although Eq.~(\ref{eqKappa}) determines mode structure in a waveguide with arbitrary cross-section, here we consider a cylindrical waveguide to illustrate our approach. In this geometry the structure of $TM$ modes is described by $E_z(r,\phi,z)\propto J_m(\kappa_{TM}^0 r){\rm e}^{im\phi+ik_z z}$. Similar considerations relate $TE$ waves to $H_z(r,\phi,z)\propto J_m(\kappa_{TE}^0 r){\rm e}^{im\phi+ik_z z}$. The structural parameters are:
\begin{equation}
\label{eqnKappa}
\kappa_{(TM|TE)}^0=\frac{X_{(TM|TE)}}{R}
\end{equation}
where $R$ is a waveguide radius and $X$ is given by $J_m(X_{TM})=0$ for {TM} waves $J_m^{\prime}(X_{TE})=0$ for $TE$ waves respectively. 

The dispersion equation Eq.(\ref{eqnDEq}) is fundamentally similar to the dispersion of a plane wave in isotropic material. Indeed, the combination $n=k_z c/\omega=\pm\sqrt{\epsilon\nu}$, which plays a role of the effective index of refraction, is combined from two (mode-dependent) scalar quantities, $\epsilon$ and $\nu$. The mode propagation requires both propagation constants to be of the same sign. 

While the parameter $\epsilon$ depends solely on the dielectric properties of the core material, the propagation parameter $\nu$ can be controlled (through $\kappa$), by changing the waveguide (or mode) geometry. Since $\kappa$ is inversely proportional to the waveguide size [see Eq.(\ref{eqnKappa})]\cite{landauECM}, there exists a {\it cut-off radius} $R_{\rm cr}\sim\lambda/2$, corresponding to $\nu=0$ for every free-space-like mode propagating in the waveguide with isotropic dielectric core. The modes propagate in structures with $R>R_{\rm cr}$, and are reflected from thinner systems. This appearance of the cut-off radius in all dielectric waveguides can be considered as a manifestation of a diffraction limit -- it is impossible to localize a combination of plane waves to a region much smaller than the wavelength inside the material $\lambda=\lambda_0 /n$, with $\lambda_0$ being free space wavelength.


Material anisotropy in a sense opens another direction in controlling the light propagation in waveguides. Indeed, anisotropy of dielectric constant makes the $TM$-mode parameters $\epsilon$ and $\nu$ completely independent of each other ($TE$ waves are not affected by material anisotropy\cite{podolskiyPRB,funnels05}). Extremely anisotropic optical materials may be used to achieve the volume mode propagation in deep subwavelength waveguides. 

As it directly follows from Eqs.(\ref{eqnDEq},\ref{eqEpsNu}), when $(\epsilon_\perp>0,\;\epsilon_\|<0)$, the parameter $\nu$ is positive regardless the size of the system. Thus, the cut-off radius does not exist at all: the decrease of the waveguide radius is accompanied by the decrease of the internal wavelength of the mode $2\pi/k_z \propto R$, in a sense ``postponing'' the diffraction limit in the system. 

\begin{figure}
\centerline{\includegraphics[width=10cm]{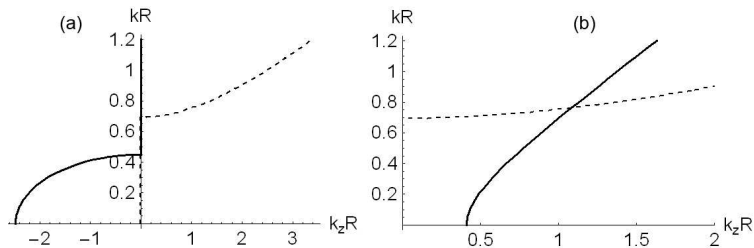}}
\caption{Dashed lines: dispersion relation of $TM_{01}$ mode in a waveguide with Si core; $R=80 nm$. Solid lines: dispersion relation of the $TM_{01}$ mode in the waveguide with anisotropic core [material dispersion is neglected]; $k=\omega/c$. Material parameters are those of Si-Ag composite described in the text for $\lambda_0=1.2 \mu m$ (a) and $\lambda_0=500nm$ (b). The former case corresponds to a negative refraction $(\epsilon_\perp<0,\;\epsilon_\|>0)$, while the latter describes positive refraction case $(\epsilon_\perp>0,\;\epsilon_\|<0)$. Positive group velocity is assumed. See text for details}
\label{figDispEq}
\end{figure}

The case of opposite anisotropy ($\epsilon_\perp<0,\; \epsilon_\|>0$), is of a special interest. The mode propagation is now possible only when $\nu<0$, which in turn {\it requires} the waveguide cross-section to be extremely small. Furthermore, causality arguments now require the phase velocity of a propagating mode to be {\it negative} (Fig.\ref{figDispEq})\cite{podolskiyPRB,funnels05}. In a sense, such a waveguide is a complete antipode of a conventional dielectric fiber, in terms of phase propagation, as well as in terms of cut-off radius. 

\begin{figure}
\centerline{\includegraphics[width=10cm]{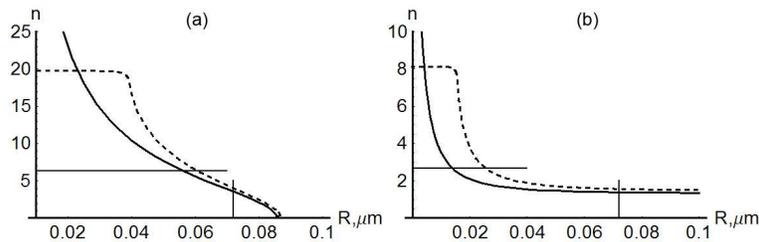}}
\caption{Effect of material inhomogeneity on the mode propagation in the waveguides with homogeneous anisotropic cores (solid lines) and in waveguides with nanolayered composite cores (dashed lines). Material parameters correspond to $TM_{11}$ mode propagation in Si-Ag system, $\lambda_0=1.2\mu m$ (a) and $TM_{01}$ mode propagation in Si-Ag system, $\lambda_0=500 nm$ (b). Horizontal and vertical lines correspond to $k_z (a+b)=1$ and to $\kappa_{TM} (a+b)=1$ respectively.
}
\label{figKz}
\end{figure}

Fig.\ref{figKz} shows the dependence of refractive index on the waveguide radius. It is clearly seen that the ``subdiffractional'' light propagation indeed follows from the reduction of the internal wavelength [increase of effective $n$] in thinner waveguides. The figure also compares the effective index of refraction in homogeneous anisotropic material to the one in anisotropic meta-material composed from a number of thin metal-dielectric layers as described below.  

\section{Material inhomogeneity and subwavelength light propagation}

Confinement of the propagating waves to deep subwavelength areas requires extremely strong anisotropy of the dielectric core. While a large number of anisotropic materials exist in nature, only a few of them exhibit the required strong anisotropy\cite{podolskiyJMO}. In this section we consider the perspectives of using nanostructured composites, known as meta-materials, as strongly anisotropic dielectric cores. Some examples of these systems include nanolayer and nanowire structure\cite{podolskiyJOSA,elserWires}. Here we consider the case when strong anisotropy is achieved in an array of metal-dielectric nanolayers, shown in Fig.\ref{figConfig}b. 

To fabricate these structures, one may use standard CVD, MOCVD, or e-beam writing techniques to build a nanolayer composite with the total height equal to the length of the anisotropic fiber, followed by ion-beam or chemical etching to ``shape'' the cylindrical or conical waveguide. E-beam writing or self-assembling techniques could be also used to directly fabricate the fiber with nanolayer core. 

When the characteristic thickness of each individual layer becomes much smaller than the wavelength, the properties of the metal-dielectric composite are well-described by effective-medium theory (EMT)\cite{podolskiyJOSA}:
\begin{eqnarray}
\label{eqnEMT}
\epsilon_\perp=\frac{a_d \epsilon_d+a_m \epsilon_m}{a_d+a_m}
\\  \epsilon_\| =\frac{(a_d+a_m)\epsilon_d \epsilon_m}{a_d \epsilon_m+a_m \epsilon_d}, \nonumber
\end{eqnarray}
where $a_d$, $a_m$ and $\epsilon_d>0$, $\epsilon_m<0$ are thicknesses and permittivities of dielectric and metal layers respectively. 

As an example, in this work we assume that the core material is composed from $15-nm$-thick layers of Ag (modeled using Drude approach) and Si ($\epsilon_{\rm Si}=12$). As it directly follows from Eq.(\ref{eqnEMT}), this system works as right-handed ($\epsilon_{\|}=-59.48+2.78 i,\;\epsilon_{\perp}=1.72+0.06 i $) for $\lambda_0=500\;nm$, and as left-handed ($\epsilon_{\|}= 28.72 + 0.12 i,\;\epsilon_{\perp}= -30.51 + 0.77 i $) for $\lambda_0=1.2\;\mu m$. The mode behavior in these systems is illustrated in Fig.~\ref{figKz}. Two approaches are used to calculate the propagation constant of the mode in each system: EMT [Eq.(\ref{eqnEMT})], and analytic solution of light transmission though a $1D$ periodic layer array\cite{yariv,funnels05}. It is seen that the predictions of both techniques are almost identical for the thicker waveguides, but strongly disagree for thinner systems. In fact, inhomogeneous microstructure of the waveguide core introduces the cut-off radius in anisotropy-based systems. While the appearance of such a critical radius may seem similar to the one in ``conventional'' dielectric fibers, the two have fundamentally different origins. In homogeneous systems the wave propagation becomes impossible when one tries to confine the propagating wave to the spatial area smaller than the wavelength. In metamaterial-based structures, on the other hand, the wavelength ``self-adjusts'' to the waveguide radius. The mode cut-off in this case has its origin in the break-up of the EMT when the internal field variation scale [$\Lambda=\min(2\pi/k_z, 2\pi/\kappa)$] becomes comparable to an inhomogeneity scale [$\Delta\simeq (a_d+a_m)$]\cite{elserSurf}. 

We note that while thinner layers may in principle allow unrestricted confinement of free-space light, in reality such a confinement will be limited by finite-size corrections to the material properties of the layers (spatial dispersion, Landau damping\cite{stockmanNano,landauPK}). For metals, the minimum layer thickness can be estimated using
\begin{equation}
a_m^{\min}\approx \lambda_0\frac{v_f}{c}\sim\frac{\lambda_0}{100}, 
\end{equation}
with $v_f$ being Fermi velocity\cite{landauPK}. 

The range of wave propagation in metamaterial-core waveguides is affected not only by spatial dispersion, but also by implicit material absorption. For the systems that can be successfully treated with EMT the field attenuation given by the imaginary part of the propagation constant $k_z$ depends on waveguide geometry along with material properties [see Eq.(\ref{eqnDEq})]. For $TM_{01}$ modes in Ag-based systems with $R=80 nm$ considered here the attenuation is dominated by absorption inside Ag. We estimate that the intensity of the light attenuates in $e$ times on the length of $1.5 \mu m$. This attenuation, although it is acceptable for the short-range signal transfer, may be further compensated or even reversed, by implementing the gain into the ``dielectric'' component of a metamaterial \cite{funnels05}. 

\section{Effect of waveguide walls}

In this section we consider the mode dynamics in waveguides with dielectric or metallic walls. Similar to the case of perfectly conducting walls described above, the light propagation in fibers with any isotropic cladding can be related to the propagating waves with $TE$ and $TM$ polarizations. In this approach, the field is represented as a series of $TM$ and $TE$ waves with same propagating constant $k_z$ and frequency $\omega$, and the boundary conditions are used to find the effective refractive index $n=k_z c/\omega$. Note that $TE$ and $TM$ components of the mode have similar {\it but not identical} spatial structure inside the anisotropic core. Explicitly, this structure is given by $J_m(\kappa_{(TE|TM)} r)\exp(i m \phi)$ with $\kappa_{TE}^2=\epsilon_\perp \omega^2/c^2-k_z^2$, and  $\kappa_{TM}^2=\epsilon_\| (\omega^2/c^2-k_z^2/\epsilon_\perp)$. The mode structure in the cladding material is described by $K_m(\kappa_{\rm cl} r) \exp(i m \phi)$ with $\kappa_{\rm cl}^2=k_z^2-\epsilon_{\rm cl} \omega^2/c^2$, and $\epsilon_{\rm cl}$ being permittivity of the cladding. The boundary condition matching yields the following dispersion relation for a propagation constant of a mode in waveguide with anisotropic core: 
\begin{eqnarray}
\label{eqnDEqReal}
\left[J_m^\ddag(\kappa_{TE} R) + K_m^\ddag(\kappa_{\rm cl} R)\right]
\left[\epsilon_\| J_m^\ddag(\kappa_{TM} R) + \epsilon_{\rm cl} K_m^\ddag(\kappa_{\rm cl} R) \right] = 
\frac{m^2 \omega^2}{R^2 c^2} \left(\frac{\epsilon_\perp}{\kappa_{TE}^2}+\frac{\epsilon_{cl}}{\kappa_{cl}^2}\right) \left(\frac{\epsilon_\|}{\kappa_{TM}^2}+\frac{\epsilon_{cl}}{\kappa_{cl}^2}\right)
\end{eqnarray} 
where $L_m^\ddag(\kappa R)=L_m^\prime(\kappa R)/[\kappa L_m(\kappa R)]$. The two terms in the left-hand side of the equation correspond to the contributions from $TE$ and $TM$ modes respectively. As it is follows from Eq.(\ref{eqnDEqReal}) the ``pure'' $TM$ and $TE$ modes are only possible when (i)~$m=0$, or (ii)~$\epsilon_{\rm cl}\rightarrow-\infty$. The latter case corresponds to perfectly conducting metallic walls described in Section 1. Solutions of Eq.(\ref{eqnDEqReal}) can be separated into two fundamentally different groups: the ones with $\kappa_{(TE|TM)}^2>0$ describe volume modes, while the ones with $\kappa_{(TE|TM)}^2<0$ correspond to surface waves. 

It is possible to dramatically simplify Eq.(\ref{eqnDEqReal}) for the case of waveguides with metallic walls. At optical or infrared frequencies, the permittivity of metals is dominated by plasma-like response of their free electrons\cite{landauPK}. As a result, this permittivity is negative and $|\epsilon_{\rm cl}|\gg 1$. A straightforward Taylor expansion of Eq.(\ref{eqnDEqReal}) yields 
\begin{eqnarray}
\label{eqKzCorr}
k_z^{TM} \simeq k_{z}^{(0)}\left(1 - \frac{\omega \epsilon_\perp}{c k_z^{(0)^2} R}\frac{1}{\sqrt{-\epsilon_{cl}}} \right),
\\k_z^{TE}\simeq k_{z}^{(0)}\left[1 + \frac{c \kappa^{(0)^2}_{TE} J_m(\kappa^{(0)}_{TE} R)}{\omega k^{(0)^2}_z R J_m^{\prime\prime}(\kappa_{TE}^{(0)} R)} \left(1+\frac{k^{(0)^2}_z m^2 \epsilon_\|}{\epsilon_\perp \kappa^{(0)^2}_{TE} \kappa^{(0)^2}_{TM} R^2} \right)\frac{1}{\sqrt{-\epsilon_{cl}}} \right],\nonumber
\end{eqnarray}
where the superscript (0) denotes the mode parameters in a waveguide with perfectly conducting walls. Note that similar to planar waveguides\cite{podolskiyJOSA}, finite value of permittivity of the waveguide wall leads to a mode expansion into the cladding region. 

\begin{figure}
\centerline{\includegraphics[width=10cm]{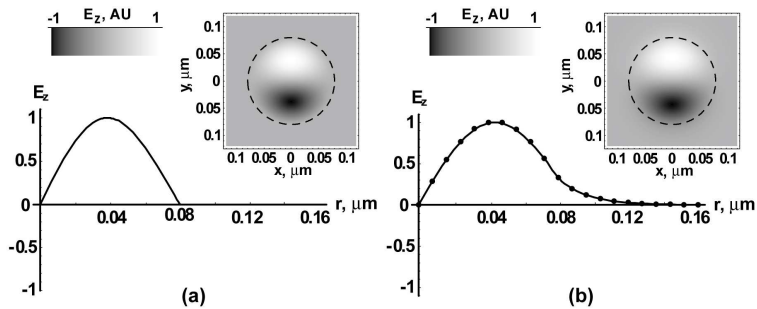}}
\caption{$TM_{11}(HE_{11})$ volume modes in waveguides with anisotropic cores. Permittivities of core material correspond to Si-Ag composite described in the text. $\lambda_0=1.2\mu m$. Panel (a) corresponds to perfectly conducting waveguide walls, panel (b) shows the mode in a waveguide with Ag cladding. Lines represent exact results [Eq.~(\ref{eqnDEqReal})]; dots correspond to perturbative Eq.(\ref{eqKzCorr})
}
\label{figModes}
\end{figure}

Besides affecting the mode propagation constant, cladding material in fiber systems also affects the mode structure. In fact, the $m\geq 1$ mode of a cylindrical waveguide with real (metal or dielectric) walls can be represented as a linear combination of $TE$ and $TM$ waves, known as $HE$ or $EH$ waves\cite{adams}. In particular, the $HE$ wave can be represented as a combination of the $TM$ mode with an admix of the $TE$ mode: $E_z\propto J_m(\kappa_{TM} r) \exp(im\phi),H_z\propto \alpha J_m(\kappa_{TE} r) \exp(im\phi)$, with the admix parameter $\alpha$ given by: 
\begin{eqnarray}
\label{eqAdmix}
\alpha = \frac{i \omega k_z m \epsilon_\| (\epsilon_\perp - \epsilon_{cl})}{c \epsilon_\perp \kappa_{cl}^2 \kappa_{TM}^2 R } 
\cdot \frac{J_m(\kappa_{TM}R)}{ J_m^\ddag(\kappa_{TE}R)+K_m^\ddag(\kappa_{\rm cl}R)}
\end{eqnarray}
Note that the effect of mode structure modification is unique to fiber-geometries and is not observed in planar waveguides. Our calculations show that for Ag walls, this admix in the $HE_{11}$ mode is below 2\%. 

As noted before, the $TE$ ($EH$) modes are (almost) unaffected by the material anisotropy. Therefore the properties of these waves are identical to the properties of $TE$ modes in waveguides with isotropic core. 

In Fig.\ref{figModes} we illustrate a propagating volume mode in the Ag-Si system described above for $\lambda_0=1.2\mu m$. We provide a comparison of the mode structure in waveguides with perfectly conducting and Ag walls. It is clearly seen that for the silver waveguide, the mode structure is well-described by the perturbative result [Eq.(\ref{eqKzCorr})]. 

\begin{figure}
\centerline{\includegraphics[width=10cm]{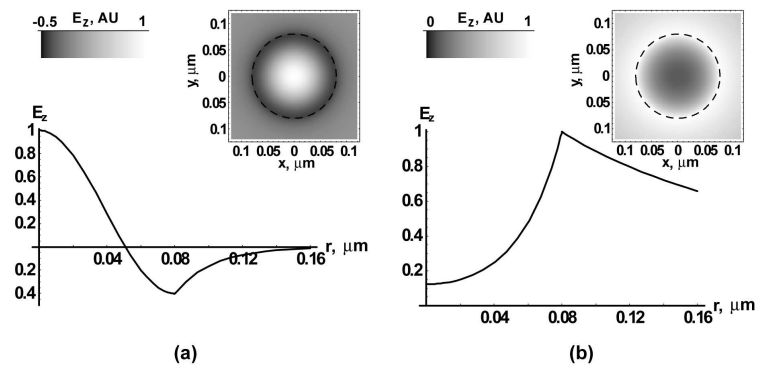}}
\caption{$TM_{02}$ mode in anisotropic fiber with free-standing anisotropic core ($\epsilon_{\rm cl}=1$) (a); SPP mode propagating at Ag-Air interface (b); $\lambda_0 = 1.2 \mu m$. Note the fundamental structural difference between volume and surface modes. Also note that volume mode has better confinement than its surface counterpart.
}
\label{figVacuum}
\end{figure}

Finally, use the exact dispersion relation [Eq.(\ref{eqnDEqReal})] to calculate the modes in the metamaterial fiber without cladding, and compare these modes to the surface {\it polariton} mode on the Ag nanowire\footnote{ In contrast to its ``right-handed'' counterpart, left-handed Ag-Si fiber described here does not support surface modes. Complete description of surface modes in strongly anisotropic systems will be presented in\cite{elserSurf}. }. Results of these simulations are shown in Fig.\ref{figVacuum}. It is clearly seen that the structure of the surface mode (localized at core-cladding boundary) is fundamentally different from the one in the volume modes. It is also seen that surface waves have weaker confinement than their volume counterparts. 

\section{Conclusions}
In conclusion, we have presented an analytical description of light propagation in waveguides with strongly anisotropic dielectric cores. We demonstrated that these systems support propagating modes even when the waveguide radius is as small as $1/15$ of free-space wavelength. We have analyzed the effect of material microstructure and waveguide cladding on mode propagation and structure, and suggested a practical realization of subwavelength structure for telecom frequencies. 

Finally, we note that our analysis can be easily generalized for the case of non-circular waveguides, for different classes of dielectric, plasmonic, or polar materials, and for different frequency ranges, from optics to IR to THz\cite{funnels05}. 

This research was partially supported by the General Research Fund, Oregon State University.

\end{document}